\documentstyle[latexsym,amssymb,epsfig,aps,floats,eqsecnum,
preprint,amsfonts]{revtex}
\def\laq{\raise 0.4ex\hbox{$<$}\kern -0.8em\lower 0.62 ex\hbox{$\sim$}}
\def\gaq{\raise 0.4ex\hbox{$>$}\kern -0.7em\lower 0.62 ex\hbox{$\sim$}}

\begin{document}
\draft
\bibliographystyle{unsrt}

\title{Magnetogenesis and the dynamics of internal dimensions}

\author{Massimo Giovannini\footnote{Electronic address: 
Massimo.Giovannini@ipt.unil.ch }}

\address{{\it Institute of Theoretical Physics, University of Lausanne}}
\address{{\it BSP-1015 Dorigny, Lausanne, Switzerland}}

\maketitle

\begin{abstract}
The dynamical evolution of internal space-like dimensions 
breaks the invariance  of the Maxwell's equations under Weyl rescaling 
of the (conformally flat) four-dimensional metric. Depending upon the number 
and upon the dynamics of internal dimensions large scale
magnetic fields can be created. 
The requirements coming from magnetogenesis together with the other 
cosmological constraints are examined under the assumption that 
the internal dimensions either grow or shrink (in conformal time)
prior to a  radiation dominated epoch. 
If the internal dimensions are growing
the magnitude of the generated  magnetic fields can seed the 
galactic dynamo mechanism. 
\end{abstract}
\vskip0.5pc
\centerline{Preprint Number: UNIL-IPT-00-16, June 2000 }
\vskip0.5pc
\noindent
\newpage
\renewcommand{\theequation}{1.\arabic{equation}}
\setcounter{equation}{0}
\section{Introduction} 
The past dynamical history of our Universe 
is not completely known and the theoretical 
understanding is often guided by consistency 
rather than by observational evidence. 
The remarkable similarity of the 
abundances of light elements in different 
galaxies leads to postulate that the 
Universe had to be dominated by radiation 
at the moment when the light elements were formed, 
namely for temperatures of approximately $0.1 $ MeV \cite{sar}. 
Prior to the moment of nucleosynthesis 
even indirect informations concerning the thermodynamical 
state of our Universe are lacking even if our knowledge 
of particle physics could give us important hints concerning 
the dynamics of the electroweak phase transition \cite{mis}.
 
The success  of big-bang nucleosynthesis (BBN) 
sets limits on  alternative cosmological scenarios. 
Departures from homogeneity \cite{hom} and isotropy \cite{iso} 
of the background 
geometry can be successfully constrained. 
Bounds on  the presence of matter--antimatter domains of various sizes
can be derived \cite{anti1,anti2}. 
BBN can also set limits on the 
dynamical evolution of internal dimensions \cite{int1,int2}.
Internal dimensions are an essential ingredient 
of  theories attempting the unification of gravitational and 
gauge interactions in a higher dimensional background like 
Kaluza-Klein theories \cite{kk}  and superstring theories \cite{ss}. 
 
Defining, respectively, $b_{BBN}$ 
and $b_0$ as the size of  the internal dimensions at the BBN 
time and at the present epoch, the maximal variation 
allowed to the internal scale factor from the BBN time 
can be expressed as $b_{BBN}/b_0 \sim 1 
+ \epsilon$ where $ |\epsilon | < 10^{-2} $ \cite{int1,int2}. 
The bounds on the  variation 
of the internal dimensions during the matter dominated epoch
are even stronger. Denoting with an over-dot the derivation with 
respect to the cosmic time coordinate, we have that 
$ |\dot{b}/b| < 10^{-9} H_0$ where 
$H_0$ is the present value of the Hubble parameter \cite{int1}.
The fact that the time evolution of internal dimensions
is so tightly constrained for temperatures lower of $1$ MeV
does not forbid that they could have been dynamical 
prior to that epoch.

An apparently unrelated observational evidence 
characterizing the present Universe is the existence 
of large scale magnetic fields \cite{first}. 
Faraday rotation measurements, Zeeman splitting estimates (when available) 
and synchrotron emission patterns indicate that 
distant galaxies seem to be endowed with a magnetic field of 
roughly the same strength of the one of the Milky Way \cite{obs,obs2} which 
is of the order of $10^{-6}$ G. 
The  similarity of the magnetic field strength in different 
galaxies led to the feeling that the origin of these large scale fields 
should somehow be connected with the cosmological consequences of 
the interplay between gravitational and gauge interactions \cite{second}.
Observations 
of magnetic fields at even larger scales (i.e  cluster , inter-cluster)
are still under debate mainly because of the 
rather problematic estimates of the 
electron density in the inter-galactic medium\cite{obs2}. 
The existence of strong magnetic fields with coherence 
scale larger than the galactic one can be of crucial importance 
for the propagation of high-energy cosmic rays.

The measured large scale (galactic) magnetic fields are assumed to 
be the result of the exponential amplification 
(due to galactic rotation) of some primeval 
seed fields \cite{second,dyn} whose typical value lies 
around $10^{-16}$--$10^{-25}$ G at the decoupling epoch. 
Several mechanisms have been invoked 
in order to explain the origin of these seeds \cite{s1,s2,s3}. 
Large scale magnetic 
fields can also have relevant physical implications in other
(related) areas of cosmology and especially in connection 
with polarization (and distortion) of the  Cosmic Microwave Background (CMB) 
\cite{pol}. 

In the present paper it is argued that there may be a relation between 
the existence of large scale magnetic fields and 
the possible occurrence of a primordial phase of the Universe 
where the internal dimensions have been dynamical. 
In short the logic goes as follows. 
Suppose that prior to BBN internal dimensions 
were evolving in time  and assume, for sake of simplicity, that 
after BBN the internal dimensions have been frozen to their present 
(constant) value. 
The evolution in time of a classical cosmological background can 
amplify a given distribution of (initially small) inhomogeneities 
of the metric and of (non conformally coupled) matter fields \cite{gri,par}. 
If the background geometry is isotropic and conformally flat 
Abelian gauge fields cannot be amplified  since their equations of 
motion are invariant under a (Weyl) rescaling of the metric tensor.
However, if the internal dimensions change in time the Weyl invariance 
of Maxwell equations is naturally broken \cite{berg} and 
electromagnetic (vacuum) fluctuations can be amplified. 

The purpose of the present investigation is to clarify if a suitable 
dynamics of the internal dimensions can produce sizable seed fields 
which could turn on the galactic dynamo mechanism and have other 
indirect effects during the evolution of the Universe. 
The plan of the paper is then the following. In Section II 
we will discuss our basic equations. In Section III we will 
perform a model independent analysis of the amplification 
of magnetic fields from monotonically evolving internal dimensions.
In Section IV some specific models of internal evolution will be studied.
Section V contains the concluding remarks

\renewcommand{\theequation}{2.\arabic{equation}}
\setcounter{equation}{0}
\section{Basic Equations} 
Consider a homogeneous and anisotropic manifold 
whose line element can be written as 
\begin{eqnarray}
ds^2 = G_{\mu\nu} dx^{\mu} dx^{\nu} = 
a^2(\eta) [ d\eta^2 - \gamma_{i j} d x^{i} dx^{j}] - b^2(\eta) \gamma_{a b} 
dy^a d y^b,
\nonumber \\
\mu,\nu = 0,..., D-1=d+n , ~~~~~~ i, j=1,..., d , ~~~~~~ 
a,b = d+1,..., d+n.
\label{metric}
\end{eqnarray}
[$\eta$ is the
conformal time coordinate related, as usual to the cosmic time $t=\int a(\eta)
d\eta$ ; $\gamma_{ij}(x)$, $\gamma_{ab}(y)$ are the metric
tensors of two maximally symmetric Euclidean 
manifolds parameterized,
respectively, by the ``internal" and the ``external" coordinates $\{x^i\}$ and
$\{y^a\}$]. 
The metric of Eq. (\ref{metric})
 describes the situation in which the $d$ external dimensions 
(evolving
with scale factor $a(\eta)$) and  the $n$ internal ones (evolving with scale
factor $b(\eta)$) are dynamically decoupled from each other \cite{gio1}. 
The results of the present investigation, however,
can be easily generalized to the case of $n$ different scale factors in the 
internal manifold.

Consider now  a pure electromagnetic fluctuation decoupled 
from the sources, representing an electromagnetic wave propagating 
in the $d$-dimensional external space such that $A_{\mu} \equiv A_{\mu}(\vec{x}, 
\eta)$, 
$A_{a} =0$. In the metric given in Eq. (\ref{metric}) the evolution 
equation of the gauge field fluctuations can be written as 
\begin{equation}
\frac{1}{\sqrt{-G}} \partial_{\mu}\biggl( \sqrt{-G} G^{\alpha\mu} 
G^{\beta\nu} F_{\alpha\beta} \biggr) =0,
\end{equation}
where $F_{\alpha\beta} = \nabla_{[\alpha}A_{\beta]}$ 
is the gauge field strength and 
$G$ is the determinant of the $D$ dimensional metric. Notice that 
if $n=0$ the space-time is isotropic and, therefore, the Maxwell's 
equations can be reduced (by trivial rescaling) to the flat space equations. 
If $n \neq 0$ we have that 
the evolution equation of the electromagnetic fluctuations propagating in the 
external $d$-dimensional manifold will receive a contribution from the internal
 dimensions which cannot be rescaled away \cite{berg}
\footnote{Notice that the electromagnetic field couples only to 
the internal dimensions through the determinant of the $D$-dimensional 
metric. In string theories, quite generically, the one-form fields are also 
coupled to the dilaton field. This case has been already analyzed in the 
context of string inspired cosmological scenarios \cite{s2}. }.
In the radiation gauge ($A_0 =0$ and $\nabla_{i} A^{i} =0$) \footnote{For a 
discussion of gauges in curved spaces see \cite{ford}}
the evolution  the vector potentials can be written as 
\begin{equation}
A_{i}'' + n {\cal F} A_{i}' - \vec{\nabla}^2 A_{i} =0, \,\,\,\,\,
 {\cal F} = \frac{b'}{b}.
\label{vec1}
\end{equation}
The vector potentials $A_{i}$ are already rescaled with respect 
to the (conformally flat) $d+1$ dimensional metric.  
In terms of the canonical normal modes of oscillations ${\cal A}_{i} = b^{n/2}
 A_{i}$ 
the previous equation can be written in a simpler form, namely 
\begin{equation}
{\cal A}_{i}''  - V(\eta) {\cal A}_{i} -\vec{\nabla}^2 {\cal A}_{i}  =0,\,\,\,
\,\, 
V(\eta) 
= \frac{n^2}{4} {\cal F}^2 + \frac{n}{2}{\cal F}'.
\label{eq1}
\end{equation}
In order to estimate the amplification of the gauge fields induced by the 
evolution 
of the internal geometry we shall consider the background metric of Eq. 
(\ref{metric})
in the case of maximally symmetric subspaces $ \gamma_{ij} = \delta_{ij}$, 
$\gamma_{a b} = \delta_{a b}$. 

Suppose now that the background geometry evolves along three different epochs. 
During the first phase (taking place for $] -\infty, -\eta_1]$) 
the evolution is truly  multidimensional. At $\eta = -\eta_1$ the 
multidimensional dynamics is continuously matched 
to  a radiation dominated phase turning, after decoupling, into 
a matter dominated regime of expansion. 
During the radiation and matter dominated stages 
the internal dimensions are fixed to their (present) constant size 
in order not 
to conflict with possible bounds arising both at the BBN time and 
during the matter-dominated epoch. 
The evolution of the external dimensions does not affect the amplification
of the gauge fields as it can be 
argued from Eq. (\ref{eq1}) :in the limit 
$n \rightarrow 0$ (i.e. conformally invariant background) Eq. (\ref{eq1}) 
reduces to the flat space equation.
 
A model of background evolution can be generically written as  
\begin{eqnarray}
&& a(\eta) = a_1 \biggl(-\frac{\eta}{\eta_1}\biggr)^{\sigma},\,\,\,\,
   b(\eta) = b_1 \biggl(-\frac{\eta}{\eta_1}\biggr)^{\lambda},\,\,\,\,\,\,
   \,\,\,\, \eta< - \eta_1,
\nonumber\\
&& a(\eta) = a_1\biggl(\frac{ \eta + 2 \eta_1}{\eta_1}\biggr),\,\,\,\,\, 
   b(\eta) = b_1,\,\,\,\,\,\,\,\,\,\,\,\,\,\,\,\,\,\,\,\,\,\,\,\,\,\,\,\,
-\eta_1 \leq \eta \leq \eta_2,
\nonumber\\
&& a(\eta) = a_1 \frac{ (\eta+ \eta_2 + 4 \eta_1)^2}{4 \eta_1 (\eta_2 +  
2 \eta_1)}, \,\,\,\,\, 
   b(\eta) = b_1,\,\,\,\,\,\,\,\,\,\,\,\,\,\,\,\,\,\,\,\,\,\,\,\,\,\,\,
\eta > \eta_2.
\label{back}
\end{eqnarray}
In the parameterization of Eq. (\ref{back})
 the internal dimensions grow (in conformal time) for $\lambda <0$ and they 
shrink for $\lambda > 0$ \footnote{To assume 
that the internal dimensions 
are constant during the radiation and matter dominated 
epoch is not strictly necessary. If 
the internal dimensions have a time variation during the radiation 
phase we must anyway impose the BBN bounds on their variation 
\cite{int1,int2}. 
The tiny variation allowed by BBN implies that $b(\eta)$ must be
 effectively constant for practical purposes. }.
By inserting this background into Eq. (\ref{eq1}) we obtain that 
for $\eta < - \eta_1$
\begin{equation}
V(\eta) = \frac{ n \lambda}{4 \eta^2}( n\lambda - 2),
\label{V}
\end{equation}
whereas $ V(\eta) \rightarrow 0$ for $ \eta > - \eta_1$.
Since $V(\eta)$ goes to zero for $\eta\rightarrow \pm \infty$ 
 we can define, in both limits,  
a Fourier expansion of ${\cal A}_{i}$ in terms of two 
distinct orthonormal sets of modes. If we  promote the classical 
fields to quantum mechanical operators in the Heisenberg 
representation we can write, for $\eta \rightarrow -\infty$
\begin{equation}
\hat{{\cal A}}^{\rm in}_{i}(\vec{x},\eta) = \int \frac{d^3 k}{(2\pi)^{3/2}}
\sum_{\alpha} 
e^{\alpha}_{i}(\vec{k})\bigl[ a_{k,\alpha} 
\phi_{k}(\eta) e^{i 
\vec{k}\cdot\vec{x}} + a_{-k,\alpha}^{\dagger} \phi_{k}^{\ast}(\eta) e^{- 
i \vec{k}\cdot\vec{x}}\bigr],
\label{ex1}
\end{equation}
where the sum runs over the physical polarizations.
For $\eta \rightarrow + \infty$ ${\cal A}_{i}$  can be expanded 
in a second orthonormal set of modes 
\begin{equation}
\hat{{\cal A}}^{\rm out}_{i}(\vec{x},\eta) 
= \int \frac{d^3 k}{(2\pi)^{3/2}}\sum_{\alpha} e^{\alpha}_{i}(\vec{k})
\bigl[ \tilde{a}_{k,\alpha} \psi_{k}(\eta) e^{i 
\vec{k}\cdot\vec{x}} + \tilde{a}_{-k,\alpha}^{\dagger} 
\psi_{k}^{\ast}(\eta) e^{- 
i \vec{k}\cdot\vec{x}}\bigr].
\label{ex2}
\end{equation}
Since both sets of modes are complete the old modes can be expressed in 
terms of the new ones 
\begin{equation}
\phi_{k}(\eta) = c_{+}(k)\,\, \psi_{k}(\eta) + c_{-}(k)\,\, \psi_{k}^{\ast}(\eta).
\end{equation}
If we insert this last equation back into Eq. (\ref{ex1}) we get
\begin{equation}
\tilde{a}_{k} = c_{+}(k)\,\,a_{k} + c_{-}(k)^{\ast}\,\,a_{-k}^{\dagger}.
\label{def}
\end{equation} 
The commutation relations within the ``in'' and ``out'' 
sets of orthonormal expansions are preserved if the two complex numbers 
$c_{+}(k) $ and $c_{-}(k)$ (the so-called Bogoliubov coefficients)
 are subjected to the constraints 
$|c_{+}(k)|^2 - |c_{-}(k)|^2 =1$ so that the (unitary) connection 
between the two asymptotic vacua is parametrized, overall, by 
three real numbers.
If the continuity of the field operators in $-\eta_1$ is imposed 
$c_{\pm}(k)$ can be determined. The continuity between the two 
asymptotic forms of the field operators is ensured provided 
the old mode functions and their first derivatives  are continuously 
matched to the new ones.
The evolution equation
satisfied by the mode functions  during the multidimensional phase
\begin{equation}
\frac{d^2 \phi_{k}}{d\eta^2} + 
\biggl[ k^2 - V(\eta)\biggr]\phi_{k} =0,
\label{phi}
\end{equation}
can be exactly solved
\begin{equation}
\phi_{k}(\eta) = \frac{ p}{ \sqrt{ 2 k}}\,\, \sqrt{ k \eta}\,\,
 H^{(2)}_{\nu}(k\eta),
\,\,\,\,\,\,\,\,\, p = \sqrt{\frac{\pi}{2}} e^{ i \frac{\pi}{4} ( 1 + 2 \nu)},
\label{s1}
\end{equation}
where $H^{(2)}_{\nu}$ is simply the Hankel function of 
second kind \cite{abr}.
For $\eta > - \eta_1 $  the solution of Eq. (\ref{eq1}) 
can be simply written as 
\begin{equation}
\phi_{k}(\eta) =  c_{+} \psi_{k}(\eta) + \psi^{\ast}_{k}(\eta),\,\,\,\,\,\,\,\,
\psi_{k}(\eta) = \frac{1}{\sqrt{2 k}} e^{ - i k(\eta + \eta_1)}.
\label{s2}
\end{equation}
The frequency mixing coefficient $c_{-}(k)$ determines the spectral number of 
produced particles in a given mode of the 
field which turns out to be $\langle n_{k} \rangle \sim |c_{-}(k)|^2$.
 By continuously  matching, 
in $\eta = - \eta_1$, the solutions 
expressed by Eqs. (\ref{s1}) and (\ref{s2}) 
the explicit form of the mixing coefficients can be obtained
\begin{equation}
c_{\mp} =  \frac{p}{2}~\bigl\{H^{(2)}_{\nu}(-k\eta_1)\bigl[
\sqrt{- k\eta_1} \mp \frac{i}{\sqrt{- k\eta_1}} 
\bigl(\nu + \frac{1}{2}\bigr)\bigr]
\pm i \sqrt{- k\eta_1} H^{(2)}_{\nu + 1}(-k\eta_1)\bigl\}.
\label{bog}
\end{equation}
The mixing coefficients determined in the approximation of a sudden change 
of the background geometry (taking place in $- \eta_1$)  lead to an 
ultraviolet divergence in the number and in the energy density 
of produced particles \cite{bir}. In fact for modes 
of comoving frequency $k^2$ larger than the height of the potential 
barrier $V(\eta)$ [see Eq. (\ref{V})] the sudden approximation 
is not adequate. The mixing coefficients, in the regime 
$k > \sqrt{V}$, should be computed 
using a smooth function interpolating between the two regimes. This 
standard analysis \cite{bir,gar} leads to 
a number of particle which is 
exponentially suppressed as $ \exp{[- q k/k_1]}$ where $k_1\sim \eta_{1}^{-1}$ 
and  $q$ is a number of order one. Thus, 
the frequency mixing of modes which never hit 
the potential barrier will be approximately neglected 
\cite{gio2} since, for these modes, 
$c_{+}(k) \sim 1$ and $c_{-}(k) \sim 0$. The modes experiencing  the 
amplification are the ones for which $k < \sqrt{V}$.  Since the 
maximal height of the barrier is set by the time scale of the transition 
(i.e. $V(\eta_1) \sim \eta_{1}^{-2}$) the results
of Eq. (\ref{bog}) can be safely estimated 
in the small argument limit namely for $|k\eta_1| <1$ of the 
corresponding Hankel functions.

Using the explicit form of the field operators the two-point  
correlation function of the magnetic field fluctuations can be computed
\begin{equation}
{\cal G}_{ij} (\vec{r}, \eta) = \langle 0_{k}|
\hat{B}_{i}(\vec{x}, \eta)
\hat{B}_{j}(\vec{x} + \vec{r},\eta) | 0_{k}\rangle = \int d^3 k 
{\cal G}_{ij}(k) e^{i \vec{k}\cdot\vec{r}}, 
\end{equation}
with
\begin{equation} 
{\cal G}_{ij}(k) = \frac{{\cal K}_{i j}}{ (2 \pi)^3}\,\,{\cal C}(\nu)\, 
k\, |k\eta_1|^{ -1 - 2 \nu},\,\,\,\, 2\, \nu= | n \lambda -1|,
\end{equation}   
and where 
\begin{eqnarray}
&& {\cal K}_{ij} = 
\sum_{\alpha}  e^{\alpha}_{i}(k) \,\,e^{\alpha}_{j}(k) = \biggl( \delta_{ij} 
- \frac{k_{i} k_{j}}{k^2}\biggr), 
\nonumber\\
&& {\cal C}(\nu) = \frac{2^{ 2 \nu -3 }}{\pi}\,\, \Gamma^2(\nu)\,\, \biggl( \frac{1}{2} 
- \nu \biggr)^2.
\label{C}
\end{eqnarray}
The magnetic energy density can be obtained by tracing over the 
physical polarizations with the result that
\begin{equation}
\rho_{B}(r) = \int \,\,\rho_{B}(k)\,\, \frac{\sin{k \,r}}{k\,r}\,\, \frac{d k}{k},
\end{equation}
where,
\begin{equation}
\rho_{B}(k) = \frac{{\cal C}(\nu)}{\pi^2}\,\,\,k^4\,\,\, | k\eta_1|^{-1 - 2 
\nu}, 
\label{sp0}
\end{equation}
is the logarithmic energy spectrum of the magnetic fluctuations \cite{s1,s2} 
expressed in terms of the comoving 
wavenumber $k$ 
which equals the physical momentum of the wave at $\eta = - \eta_1$
[we set $a(-\eta_1) \equiv a_1 = 1$]. At later times the 
momentum is given by $\omega(\eta) = k/a(\eta)$ and the 
wavelength is $2 \pi a(\eta)/k $.

For $\eta <- \eta_1$ Eq. (\ref{eq1}) 
is written in vacuum (no conductivity is present). 
As soon as the 
Universe gets into the radiation dominated phase the conductivity $\sigma_c$ 
suddenly jumps to a finite value of the order of $T_1$, namely the
temperature right after $-\eta_1$. Thus, 
the electric fields will be rapidly dissipated whereas 
the magnetic fields will not be damped, namely
\begin{equation}
{\cal B}_{i}(\eta_1 +\Delta \eta ) \sim {\cal B}_{i} (\eta_1),\,\,\,\,\,
{\cal E}_{i}(\eta_1 +\Delta \eta ) \sim e^{- \sigma_c a_1 \Delta\eta}  
{\cal E}_{i} (\eta_1).
\end{equation}
Indeed, when the conductivity jumps to finite value, 
the magnetic fields survive whereas the electric fields are dissipated 
in a time $\Delta\eta$ which is proportional to $T_1^{-1}$ \cite{noi}.
This physical situation is different from the one 
occurring in a tokamak \cite{kra} where the system is 
non-relativistic since the mass of the charge carriers 
is normally much larger than the temperature of the plasma.

In the absence of  vorticity in the bulk 
velocity field of the  plasma (i.e. if $\vec{\nabla} \times \vec{v} =0$), 
a given mode ${\cal B}_{i}(\omega)= 
B_{i}(\omega) a^2(\eta)$ of the magnetic field 
will approximately evolve as  
\begin{equation}
B_{i}(\omega,\eta) a^2(\eta) 
\sim B_i(\omega,\eta_1) a^2(\eta_1)e^{ - \frac{\omega^2}{\sigma} \eta}.
\end{equation}
where $\sigma = \sigma_c a(\eta)$ is the (curved-space) 
form of the conductivity. 
Provided  that $ \omega < \omega_{\sigma} \sim 
\sqrt{\sigma/\eta}$ the magnetic flux 
is conserved and the magnetic energy density red-shifts as 
$[a(\eta)]^{-4}$. 
Magnetic fields whose 
typical momentum is much smaller than $\omega_{\sigma}$ 
( sometimes called magnetic diffusivity scale) correspond to 
the large scales relevant for the dynamo action.

If some  primordial vorticity is present 
(i.e. $\vec{\nabla} \times \vec{v} \neq 0$) the evolution 
of the magnetic fields cannot be disentangled from the 
evolution of the bulk velocity field \cite{obs,gio3}. 
The parity breaking (helical) velocity field
could be the result  of some dynamically developed turbulence 
\cite{oles}, of some specific initial conditions \cite{second}, 
 or of the gravitational collapse \cite{obs}. 
Indeed, the galaxy is formed with some 
typical rotation period. The rotation of the galaxy will switch on 
a further term in the evolution equations of the magnetic fields: the 
so-called  dynamo term describing the amplification 
caused in the magnetic flux by the action of a topologically 
non-trivial bulk velocity field
(i.e. $ \langle \vec{v} \cdot \vec{\nabla} \times \vec{v}\rangle \neq 0$ )
\cite{obs,obs2}. In the presence of the dynamo action the seed field 
possibly generated by the variation of the internal manifold
 will be exponentially amplified. The number of 
e-folds characterizing the amplification is of the order of the number of 
rotations performed by the galaxy since its origin.
In order to allow a successful dynamo action (and a successful genesis 
of the large scale magnetic field) the original seed field  
(exponentially amplified later on)  
should have a specific value. The problem is to check 
 if the seed fields generated from the evolution of the 
interval dimensions will be strong enough to be amplified to the 
presently observed value. This analysis  will be one of the 
purposes of the following section.

\renewcommand{\theequation}{2.\arabic{equation}}
\setcounter{equation}{0}
\section{Magnetogenesis constraints} 
The amplified gauge fields should be compared with various constraints 
coming from general cosmological considerations (critical energy density, 
possible anisotropies induced in the cosmic microwave background) 
and from magnetogenesis. If   the produced fields
are strong enough to seed the galactic dynamo mechanism \cite{obs,obs2} 
they should also be  compatible with the physics described 
by the standard model of cosmological evolution.

Suppose that the time evolution of the internal dimensions is 
monotonic  in conformal time. In this case the internal dimensions 
will  either 
grow or shrink. If the internal dimensions 
shrink (to Planckian or quasi-Planckian size) they are 
very small today (of the order of $10^{-33}$--$10^{-32}$ cm). 
If they expand they could lead, today, 
to a smaller (effective) four-dimensional Planck mass \cite{dim} and to 
a larger value of  $b_1$ which 
could  lie between $10^{-33}$ cm and $10^{-4}$ cm.

Some theoretical models which could describe 
the multidimensional evolution are discussed in Section IV.
Here a general analysis will be given. 
The discussion is divided into two parts. 
In the first part $n$ internal dimensions growing in 
conformal time will be considered. This cooresponds, within 
the parametrization of Eq. (\ref{back}), to the case $\lambda <0$.
In the second part 
internal $n$ internal dimensions shrinking in conformal 
time will be considered. This corresponds to $\lambda >0$ in 
Eq. (\ref{back}).

\subsection{ Internal dimensions growing in conformal time}
According 
to Eq. (\ref{back}) the growth of the internal dimensions 
stops in $\eta = -\eta_1$ when the radiation dominated phase 
suddenly begins. The curvature at which the transition 
to the radiation dominated phase occur is given by 
$H_1 = T_1^2/[M_{4} ( b_1 M_{4})^{n/2}]$ where $T_1$ is the 
temperature at $-\eta_1$,$M_{4}$ is the four dimensional 
Planck scale and $H_1$ is the Hubble factor $H= \dot{a}/a$ in 
$-\eta_1$. The maximal temperature of the 
Universe cannot be larger, in this context, than 
$M_{4}$ so  $T_1 \leq M_{4}$ will be assumed.
The  present (physical) frequencies relevant for  the problem 
can be obtained using the background evolution 
during the radiation and matter dominated epochs.  
After $\eta= - \eta_1$, the evolution of the 
geometry is adiabatic. Thus, the scale corresponding today 
to $1/\eta_1$ is given by $\omega_1 \sim 10^{11} (M_{4}/M_{P})$ Hz 
(where $M_{P}\sim 1.22\times 10^{19} $ GeV). The 
decoupling frequency (corresponding to the transition from 
radiation to matter) is $\omega_{\rm dec} \sim
10^{-16}$ Hz. 

As discussed in the previous Section, thanks 
to magnetic flux conservation, the ratio between the logarithmic 
energy spectrum and  the radiation energy density is approximately 
constant 
\cite{s1,s2} and, for $T_1\sim M_{4}$, it is given by 
\begin{equation}
r(\omega) = \biggl( \frac{M_{4}}{M_{P}}\biggr)^{4} 
\frac{{\cal C}(\nu)}{\pi^2}
\biggl( \frac{\omega}{\omega_1}\biggr)^{3 - 2\nu},  
\label{r1}
\end{equation}
where, for simplicity,  we set  $\lambda = - \beta$ with $\beta >0$ so that 
$2\nu = | n\lambda -1| \equiv n\beta  + 1$. Notice that 
the effective number of relativistic species at $\eta_1$ 
has been included in the definition of $M_{4}$.
Eq. (\ref{r1}) can be rewritten  as  
\begin{equation}
r(\omega) =\frac{1}{\rho_{\gamma}} \frac{d \rho_{B}}{ d \ln{\omega}}=
 \biggl( \frac{M_4}{M_{P}}\biggr)^{ 2 + n\beta} \frac{2^{n\beta - 4}}{\pi^3}
 \biggl(\frac{ \omega }{10^{11} {\rm Hz}}\biggr)^{ 2 - n \beta} 
\Gamma^2\biggl(\frac{ n\beta + 1}{2}\biggr) |n\beta|^2,
\label{sp1}
\end{equation}
where $\rho_{\gamma} $ is the energy density in radiation.
If $n\beta < 2$,  the logarithmic 
energy spectrum  grows in  frequency. 
The critical energy density constraint is then implemented by requiring that 
$r(\omega_1)< 1$, namely by requiring that 
the magnetic energy density (evolving according to flux conservation) 
is smaller than the radiation density. This requirement
automatically guarantees, because of the growth in frequency  of the logarithmic 
energy spectrum, 
 that $r(\omega)< 1$ also for all $\omega < \omega_1$, 
and in particular, for $\omega_{\rm dec}$. 

If $n \beta > 2 $  the spectrum decreases in frequency. 
In order to insure 
the  compatibility with present large scale 
observations at low frequencies the spectrum 
should not induce too much anisotropy in cosmic microwave background (CMB). 
This condition 
is enforced if  $r(\omega_{\rm dec}) < 10^{-10}$. 
If $r(\omega_{\rm dec}) < 10^{-10}$ and $r(\omega)$ decreases 
in frequency, then, $r(\omega) <1$ for any $\omega_{\rm dec} < \omega
< \omega_{1}$.

\begin{figure}
\begin{center}
\begin{tabular}{|c|c|}
      \hline
      \hbox{\epsfxsize = 7 cm  \epsffile{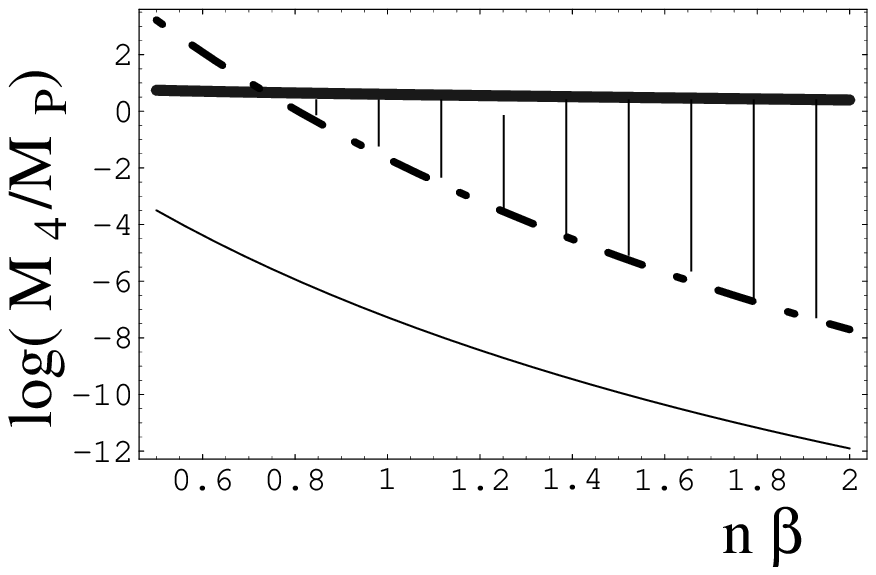}} &
      \hbox{\epsfxsize = 7 cm  \epsffile{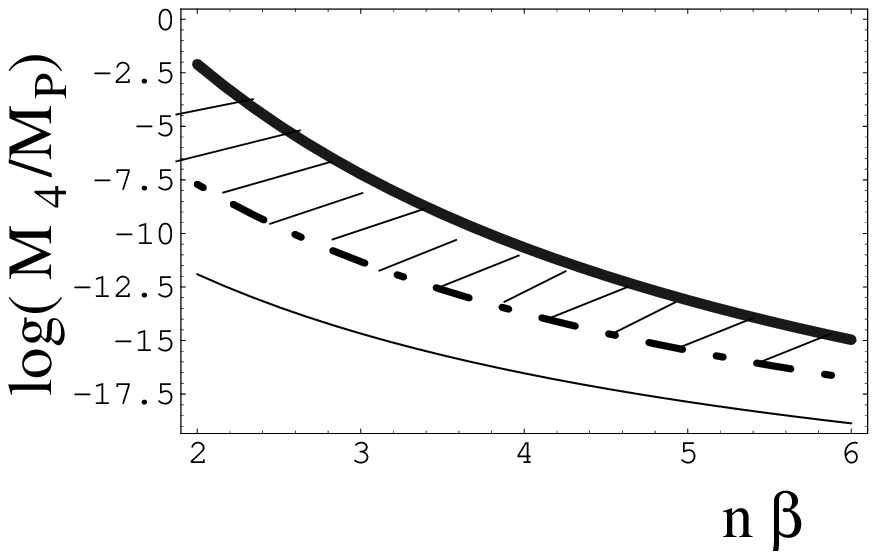}} \\
      \hline
\end{tabular}
\end{center}
\caption[a]{ The allowed region is illustrated
for the case of internal dimensions growing in conformal time 
(i.e. $\beta >0$). 
With the thick (full) line  
the requirements imposed by cosmological 
constraints are reported. The constraints set by 
magnetogenesis are illustrated
in the case of a dynamo rate $\Gamma^{-1} = 0.5$ Gyr (dot-dashed line),
and in the (less conservative) case $\Gamma^{-1} = 0.3 $ Gyr (thin line).}
\label{f1}
\end{figure}

Thus, compatibility with the observed 
features of our present Universe demands
\begin{eqnarray}
&& r(\omega_1) < 1,\,\,\,\,\, {\rm for }\,\,\, n\beta < 2,
\nonumber\\
&& r(\omega_{\rm dec}) < 10^{-10},\,\,\,\,\, {\rm for } \,\,\,\, n\beta > 2.
\label{con1}
\end{eqnarray}
Eqs. (\ref{con1}) imply, respectively \footnote{ 
Recall that $\ln$ denotes the Neperian logarithm and  $\log$ denotes 
the logarithm in ten basis.},
\begin{eqnarray}
&& 4 \log{\biggl(\frac{M_{4}}{M_{P}}\biggr)} <
 \log{\biggl[\frac{\pi^3}{(n\beta)^2}\biggr]} - ( n\beta - 4)\log{2} - 
2\log{\biggl{[\Gamma\biggl(\frac{n\beta + 1}{2}\biggr)\biggr]}},\,\,\ n\beta <2
\label{crita}\\
&& (n\beta + 2) \log{\biggl(\frac{M_{4}}{M_{P}}\biggr)} < - 8.55   
+ 27 ( 2 - n\beta) - 2 \log{n \beta}
\nonumber\\
&&- 2 \log{\biggl[\Gamma\biggl(
 \frac{ n\beta + 1}{2}\biggr)\biggr]} 
 + ( 4 - n \beta) \log{2},\,\,\,\,\,\,\,\,\, n\beta >2.
\label{critb}
\end{eqnarray} 
Magnetogenesis demands that strong seed fields 
should be produced. Such a demand \cite{obs,s2} translates 
into the following equations
\begin{equation}
r(\omega_{G}) \geq 3.7\times 10^{-33},\,\,\, \omega_{G} \sim 10^{-14} \,\, 
{\rm Hz}
\label{d1}
\end{equation}
where $ \omega_{G}^{-1} \sim 1$ Mpc. The condition expressed in 
 Eq. (\ref{d1})  mildly depends upon the cosmological 
parameters. Eq. (\ref{d1}) assumes $ h_0 = 0.65$, $\Omega_{\rm m} = 0.3$ 
and $\Omega_{\lambda} =0.7$ and it corresponds to a seed field 
of $2.5\times10^{-16}$ G at the decoupling epoch. 
The seed field generated through 
our mechanism will be exponentially amplified by the galactic 
rotation. The amplification induced by the differential 
rotation of the primeval galaxy through the dynamo 
mechanism goes roughly as $e^{ \Gamma t}$ where $\Gamma$ is the 
dynamo amplification rate and $t$ is the galactic age (roughly $10$ Gyrs)
 \cite{obs,obs2}.
In Eq. (\ref{d1}) it is assumed that the dynamo amplification rate 
is of the order of $\Gamma^{-1} \sim 0.5$ Gyr. 
If the dynamo amplification rate is 
 (more or less artificially) increased \cite{dyn3}
the initial seed can be even smaller that the one assumed in 
Eq. (\ref{d1}). 
For instance if we take $\Gamma^{-1} \sim 0.3$ the 
seed field is required to satisfy $ r(\omega_{G}) > 6.07\times 10^{-50}$ 
\cite{dyn3}.
According to Eq. (\ref{d1}), Eq. (\ref{sp1}) implies
\begin{eqnarray}
&&(2 + n\beta) \log{\biggl(\frac{M_{4}}{M_{P}}\biggr)} \geq 
\nonumber\\
&&- 30.92 +
25( 2 - n\beta) - 2\log{\biggl[\Gamma\biggl(\frac{n\beta 
+ 1}{2}\biggr)\biggr]} - 
2 \log{n\beta} - ( n\beta - 4) \log{2},
\label{magn}
\end{eqnarray}
The constraints imposed by 
Eqs. (\ref{crita})--(\ref{critb}) and by Eq. (\ref{magn}) 
entail a region in the parameter space 
of the model where the cosmological constraints can be 
safely satisfied and the magnetogenesis requirements 
are met.
The parameter space  (defined by $n\beta$ and by $M_{4}/M_{P}$) 
is illustrated in Fig. \ref{f1}. 
The shaded area corresponds to the allowed region.
In the shaded area of Fig. \ref{f1} (left plot) magnetic fields are 
produced with growing frequency spectrum. In the 
shaded area of the right plot of Fig. \ref{f1} magnetic fields 
are produced with decreasing frequency spectrum.
The theoretical bias would point, a priori, towards 
increasing frequency spectra. In fact in this case 
the two-point functions are decreasing at large distance scales.
However, in the present analysis all the possibilities should be 
borne in mind.
The full (thin) line appearing in both plots of Fig. \ref{f1} 
denotes a less conservative magnetogenesis 
requirement and it corresponds to larger  a dynamo amplification 
rate. The thin curve can be obtained by requiring, from Eq. 
(\ref{sp1}), $r(\omega_{G}) > 6.07 \times10^{-50}$ which corresponds 
to $\Gamma^{-1} \sim 0.5$ Gyr.
According to Fig. \ref{f1} reasonable seeds are produced 
 provided $n\beta$ is larger than $0.6$ . The four dimensional 
Planck mass should be  small in units of the fundamental Planck mass. 
The parameter space seems to 
select typical values of $M_{4}$ between $10^{13}$ Tev and $10^{3}$ TeV.

These conclusions were reached in the case $T_1 \simeq M_{4}$. If
$T_1 \ll M_{4}$ similar conclusions can be reached 
by following the same steps. The conclusion is that 
the intercepts between the thick and dot-dashed curves 
are slightly shifted to the left. To give some 
numerical value, assume, for instance,  
$T_1 \sim 10^{-4} M_{4}$.  
Then we would get that the intercept gets to 
$n\beta \simeq 0.8$. 

In order to avoid confusion 
we would like to stress that in spite of the fact 
that, in our case, $M_{4}/M_{P} \ll 1$ the physical size 
of the internal dimensions is always much smaller than $0.1$ mm 
corresponding to a $M_4 \sim $ TeV. Therefore, further 
constraints coming from submillimiter tests of the Newton law 
are not directly relevant to our present discussion.

To complete the analysis it is 
worth mentioning that magnetic fields can be 
directly constrained from BBN. These bounds 
are qualitatively  different from the ones previously quoted and
coming, alternatively, from homogeneity \cite{hom} and isotropy 
\cite{iso} of the background geometry at the BBN time.
As elaborated in slightly different frameworks
 through the years \cite{bbn}, magnetic fields 
possibly present at the BBN epoch could have a twofold effect. 
On one hand they could 
enhance the rate of reactions (with an effect
proportional to $\alpha \rho_{B}$ where $\alpha \sim 1/137$) 
and, on the other hand they could 
artificially increase the expansion rate (with an effect 
proportional to $\rho_{B}$). It turns out that 
the latter effect is probably the most relevant. 
 In order to prevent the Universe 
from expanding too fast at the BBN epoch 
$\rho_{B} < 0.27 \rho_{\nu}$  where  $\rho_{\nu}$ is the 
energy density contributed by the standard three neutrinos for $T< 1$ MeV.
This bound can be imposed also to the  spectra discussed in the present 
Section. Taking into account that the  horizon at the 
BBN epoch (assuming a typical temperature of $0.1 $ MeV) corresponds 
to a present frequency of $\omega_{{\rm NS}}\sim 2.28\times 10^{-12}$ Hz, 
the BBN bound implies in the case of the spectra of Eq. (\ref{sp1}) 
\begin{eqnarray}
&& ( 2 + n \beta) \log{\biggl(\frac{M_{4}}{M_{P}}\biggr)} <
 0.19 + ( 4 - n\beta)\log{2} + 22.6\,(2 - n\beta)
\nonumber\\
&&  -
2 \log{\biggl[\Gamma\biggl(\frac{n \beta + 1}{2}\biggr)\biggr]} - 2 \log{n\beta}
\end{eqnarray}
This constraint is  qualitatively different from the ones  already 
discussed and that is why it is important to mention it. 
However, from the  quantitative point of view it is roughly equivalent to 
the critical density constraint. If $n\beta >2$ this bound 
is always satisfied. If $n\beta <2$ this bounds is a bit more 
constraining than the critical density limit
(by a tiny amount) in the region $1.4 <n\beta < 2$. 
Within the numerical accuracy of the present analysis it is not crucial.

Another interesting moment in the life of the 
Universe where the generated magnetic fields could have 
an impact is the electroweak epoch \cite{ew}. The present 
frequency corresponding at the electroweak 
horizon is of the order of $10^{-5}$ Hz. Fields coherent 
on that scale can be certainly produced within
the present mechanism but they are not constrained 
by the electroweak physics. 

\subsection{Internal dimensions shrinking in conformal time}
If $\lambda >0$ (internal dimensions shrinking in conformal time), 
then
$2 \nu = |n \lambda -1|$. 
Therefore, the logarithmic energy spectrum has 
a twofold form
\begin{eqnarray}
&& r(\omega) = \biggl( \frac{H_{1}}{M_{P}}\biggr)^2 
\biggl(\frac{\omega}{\omega_1}\biggr)^{4 - n \lambda} 
\frac{ 2^{ n\lambda - 6}}{\pi^3} 
( 2 - n\lambda)^2 \Gamma^2
\biggr(\frac{n\lambda -1}{2}\biggl),\,\,\, {\rm for }\,\,\, n\lambda > 1
\nonumber\\
&& r(\omega) = \biggl(\frac{H_1}{M_{P}}\biggr)^2 
\biggl(\frac{\omega}{\omega_1}\biggr)^{2 + n\lambda}
\frac{ 2^{- 2 - n \lambda}}{\pi^3} 
(n\lambda)^2 \Gamma^2\biggl(\frac{1 - n \lambda}{2}\biggr),\,\,\,
 {\rm for }\,\,\, 
n\lambda <1
\label{cont}
\end{eqnarray}
If $ n\lambda< 4$ the spectrum is always increasing in frequency.
 On the contrary 
for $n\lambda > 4$ the spectrum decreases and it is peaked in the infra-red. 
In order to implement the 
critical density bound  for $n \lambda < 4$,  
$r(\omega_1) < 1$ should be demanded.
For $ n\lambda > 4$, $r(\omega_{\rm dec}) < 10^{-10}$ should be required 
in order to guarantee the compatibility of our toy model with the CMB 
anisotropies. 
Then, using Eq. (\ref{cont}), the critical density bound demands
\begin{eqnarray}
&& 2 \log{\biggl( \frac{H_1}{M_{P}}\biggr)} \leq ( 2 + n\lambda) \log{2} + 
\log{\pi^3}
- 2 \log{n\lambda} - 2 \log{\biggl[\Gamma\biggl(\frac{1 
- n\lambda}{2}\biggr)\biggr]}, \,\,\,\,\, n\lambda <1.
\\
&& 2 \log{\biggl( \frac{H_1}{M_{P}}\biggr)} \leq ( 6 - n \lambda) 
+ \log{\pi^3} 
\nonumber\\
&&- 2 \log{(n\lambda -2)} - 
2 \log{\biggl[\Gamma\biggl(\frac{n \lambda -1}{2}\biggr)\biggr]}, 
\,\,\,\,\ 1<n\lambda < 4.
\end{eqnarray}
In this case the constraints are expressed using 
curvature  (rather than 
temperature ) scales. This is because
 $H_1 \sim 1/(a_1 \, \eta_1)$ cannot be much smaller than $M_{P}$ 
In this language the (present) maximal 
frequency 
turns out to be $\omega_1 = 10^{11} \,\sqrt{H_1/M_{P}}$  Hz.
Using Eq. (\ref{cont}) in the case $n \lambda > 4$ the CMB bound becomes 
\begin{eqnarray}
&& n \lambda \log{ \biggl(\frac{H_1}{M_{P}}\biggr)} \leq
\nonumber\\
&& -17.01 + 54 ( 4 - n \lambda) 
- 2 ( n\lambda -6) \log{2}  -2 \log{(2- n\lambda)^2} - 
4 \log{\biggl[
\Gamma\biggl(\frac{n \lambda - 1}{2}\biggr)\biggr]}
\end{eqnarray}
Finally  the magnetogenesis requirements [already introduced in Eq. 
(\ref{d1})] impose
\begin{eqnarray}
&& (2 - n\lambda)  \log{\biggl( \frac{H_1}{M_{P}}\biggr)}\geq - 61.88 
+ 50 ( 2 + n \lambda)
\nonumber\\
&& 
+ 2 (2 + n\lambda) \log{2} - 4 \log{n\lambda} - 4 \log{\biggl[
\Gamma\biggl(\frac{1-n \lambda }{2}\biggr)\biggr]},\,\,\,\, n\lambda <1 
\\
&& n\lambda  \log{\biggl( \frac{H_1}{M_{P}}\biggr)} \geq- 61.88 
+ 50(4 - n\lambda)
\nonumber\\
&& - 
2 ( n\lambda - 6) \log{2} - 4 \log{( n\lambda -2 )} - 4 \log{\biggl[
\Gamma\biggl(\frac{n \lambda -1 }{2}\biggr)\biggr]},\,\,\,\, n\lambda > 1 
\label{d2}
\end{eqnarray}
\begin{figure}
\begin{center}
\begin{tabular}{|c|c|}
      \hline
      \hbox{\epsfxsize = 7 cm  \epsffile{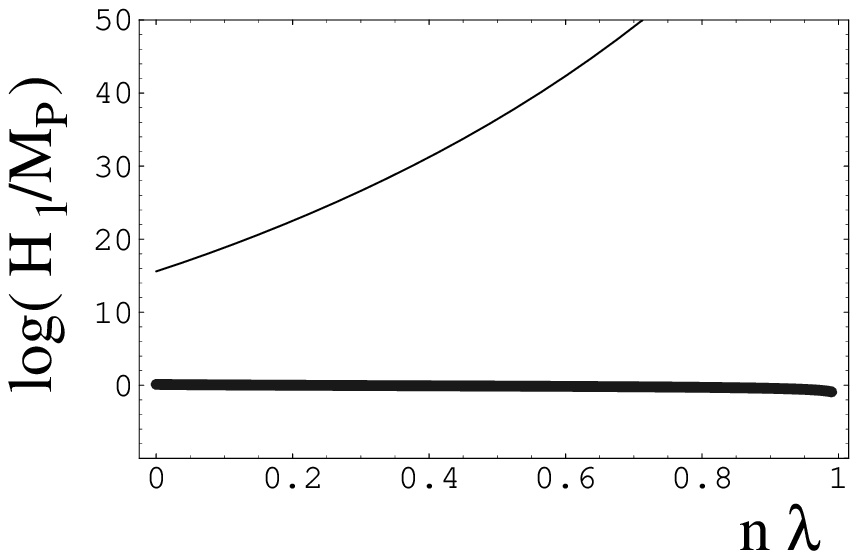}} &
      \hbox{\epsfxsize = 7 cm  \epsffile{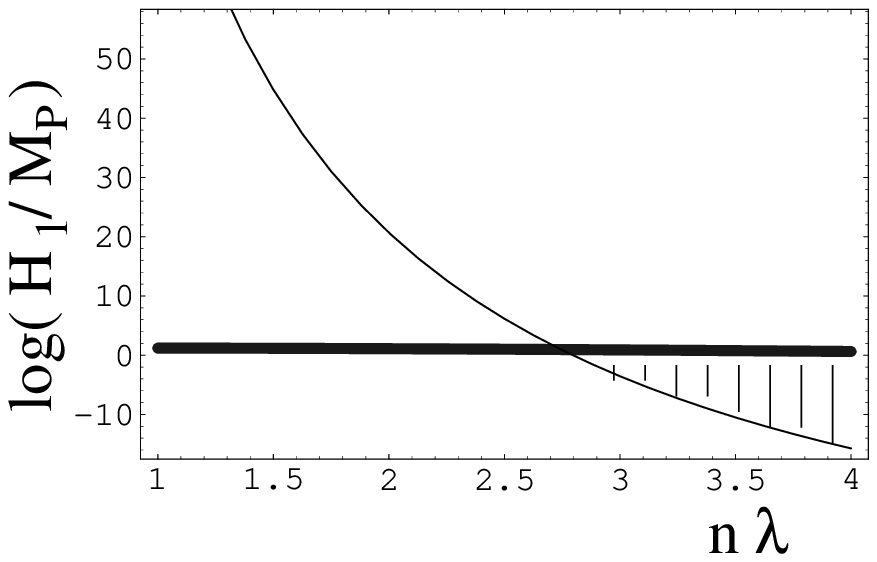}} \\
      \hline
\end{tabular}
\end{center}
\caption{The allowed region in the case of $\lambda >0$ is 
illustrated. The magnetogenesis requirements following from 
Eqs. (\ref{d2}) are reported with full (thin) lines. 
The thick lines denote the cosmological 
bounds. For $n\lambda <1$ magnetogenesis 
cannot take place since there is no intercept between 
the cosmological bounds and the magnetogenesis 
region. If $n\lambda >1$  magnetogeneis 
demands $n\lambda > 2.7$.}
\label{f2}
\end{figure}
Fig. \ref{f2} shows that, once the cosmological bounds are imposed,
 magnetogenesis is possible only of the parameter space implying that
$n\lambda >1$. If $n \lambda < 1$ magnetogenesis 
is excluded unless the produced magnetic fields 
over-close the Universe. This possibility 
must be rejected. 

\renewcommand{\theequation}{3.\arabic{equation}}
\setcounter{equation}{0}
\section{Tailoring the model of internal evolution} 
In the previous Sections it has 
been established that large scale magnetic fields can be 
produced either if $n\beta \geq 0.6$ (if the internal dimensions 
grow in conformal time) or if $n\lambda > 2.7$ (if the 
internal dimensions shrink in conformal time)
. Some toy model of internal 
evolution will now be analyzed. The purpose 
of this analysis is to show that the physical requirements 
coming from magnetogenesis and from other cosmological considerations
can be indeed realized in specific models of dynamical evolution 
of a $D$-dimensional background geometry. 

\subsection{General considerations}
Consider the $D$-dimensional action for a scalar field, minimally
coupled to gravity:  
\begin{equation}
S= S_{g} + S_{m}= - \int d^D x \sqrt{-g}R +
\int d^D x \sqrt{-g}
\left[\frac{1}{2}g^{\alpha\beta}\partial_{\alpha}\varphi\partial_{\beta}\varphi
- V(\varphi)\right] ,        
\label{action}
\end{equation}
where we work, for simplicity, in Planck units.   
We shall consider a homogeneous, Bianchi-type I metric background,
whose spatial part is the product of two conformally flat manifolds as 
introduced in Eq. (\ref{metric}).  
For the  background given in  Eq. (\ref{metric}), 
the equations of motion obtained by varying the
action with respect to $g_{\mu\nu}$ and $\varphi$, 
\begin{equation}
R_\mu^\nu - \frac{1}{2}\delta_{\mu}^\nu R = \frac{1}{2} \left[ 
\partial_\mu\varphi\partial^\nu \varphi -
\frac{1}{2}\delta_\mu^\nu
g^{\alpha\beta}\partial_\alpha
\varphi\partial_\beta\varphi+ \delta_\mu^\nu
V(\varphi)\right] , ~~~~~~
g^{\alpha\beta}\nabla_{\alpha}\nabla_{\beta}\varphi+ \frac{\partial
V}{\partial\varphi}=0 , 
\label{ein}
\end{equation}
reduce simply to 
\begin{eqnarray}
{d(d-1)}{\cal H}^2+{n(n-1)} {\cal F}^2+ 2n d {\cal H}
{\cal F}=  \left(\frac{\varphi'^2}{2} +  a^2
V\right),~~\nonumber\\ 
2(d-1) {\cal H}'+{(d-1)(d-2)}{\cal H}^2+2n
{\cal  F}'+{n(n+1)} {\cal F}^2+ 2n (d-2){\cal H}{\cal F}
&=& \left(a^2 V -\frac{\varphi'^2}{2} \right),\nonumber \\
2(n-1){\cal F}'+ 2d {\cal H}'+{d(d-1)} {\cal H}^2+
{n(n-1)}{\cal F}^2+2 (d-1)(n-1) {\cal H}
{\cal F}
&=& \left(a^2 V -\frac{\varphi'^2}{2}\right), 
\nonumber\\
\varphi''+\left[(d-1){\cal H} +n {\cal
F}\right]\varphi'+\frac{\partial V}{\partial\varphi} =0 
,~~~~~~~~~~~~~~ 
\label{background}
\end{eqnarray}
where ${\cal H}= {(\ln{ a})}^{\prime}$, ${\cal F}= {(\ln{b})}^{\prime}$.
 
These equations are not all independent, and the scalar field 
 equation, for instance,    
can be obtained from the other Einstein equations. 
By summing and subtracting the above equations one obtains
\begin{eqnarray}
 a^2 V &=&  \left[(d-1){\cal H} + n{\cal F}\right]' +
\left[(d-1){\cal H} +n{\cal F}\right]^2 ~ ,  
\label{first}\\
\frac{1}{2} \varphi'^2 &=& -\left[(d-1){\cal H} +n{\cal F}\right]' +
 (d-1){\cal H}^2 -n {\cal F}^2 + 2 n {\cal H} {\cal F}~, 
\label{second}\\
{\cal F}' -{\cal H}' &=& -( {\cal F} - 
{\cal H}) \left[(d-1){\cal H} +n {\cal F}\right]~~~~.
\label{third}
\end{eqnarray}
Consider now the case where the external and the internal scale 
factors can be parametrized as $a(\eta) \sim (- \eta)^{\sigma}$ and 
$b(\eta)\sim (-\eta)^{\lambda}$. If $V=0$ and $ \varphi=0$ 
we have that the well known solution of this sysmtem is indeed given 
by a power law behaviour of the scale factors provided 
the exponents satisfy, according to Eqs. (\ref{first})--(\ref{third}) 
\begin{eqnarray}
&& (d -1) \sigma + n \lambda = 1,
\label{sum}\\
&& ( d-1) \sigma^2 + n \lambda^2= 1 + 2 \lambda.
\label{sq}
\end{eqnarray}
These two conditions can be also easily expressed in cosmic time . 
In this case we have to recall that $a(\eta) d\eta = dt$. Therefore 
if we parameterize the solutions in cosmic time as 
\begin{equation}
a(t) \sim (-t)^{\epsilon},\,\,\,\,\, b(t) \sim (-t)^{\zeta} 
\label{par}
\end{equation}
we will have that the exponents 
in cosmic and conformal time are simply related as 
$\epsilon = \sigma/ (\sigma + 1)$ and $ \zeta= \lambda/(\sigma + 1)$.
Using these two last relations into Eqs. (\ref{sum})--(\ref{sq}) 
 the two standard Kasner conditions are recovered, namely
 $d \epsilon + n \zeta =1$ and  $ d\epsilon^2 + n \zeta^2 =1$. 
The solution of Eqs. (\ref{sum})--(\ref{sq})  gives, for each dimension 
$n$ of the internal space, two conjugate solutions which we write
in the case $d =3$:
\begin{eqnarray}
&& \sigma_{\pm} = \mp \frac{1}{2} \biggl[ \sqrt{\frac{ 3 n}{n + 2}} 
\mp 1\biggr],
\nonumber\\
&& \lambda_{\pm} = \pm \sqrt{\frac{3}{n ( n+ 2)}}.
\end{eqnarray}
The related cosmic time exponents can be easily obtained 
\begin{eqnarray}
&& \epsilon_{\pm} = \frac{ 1 \mp \sqrt{\frac{n + 2}{3 n}}}{ 1 \mp 
3 \sqrt{\frac{ n + 2}{3n}}},
\nonumber\\
&& \zeta_{\pm} = \frac{ - 2 }{ n ( 1 \mp 3 \sqrt{\frac{n + 2}{3 n}})}.
\end{eqnarray}
In the parameterization of Eq. (\ref{par}) 
 the solution labeled by $+$ in the exponents 
of the internal manifold denote a growing solution 
whereas the solution denoted by $-$ denote 
a shrinking solution. 
In order to produce large scale magnetic fields we should require 
that, when the internal dimensions expand, 
\begin{equation}
|n \lambda_{-} | > 0.6.
\label{c1}
\end{equation}
If we now insert the explicit expression for $\lambda_{-}$ we have that 
Eq. (\ref{c1}) implies 
\begin{equation}
\sqrt{\frac{3 n}{ n + 2}} \geq 0.6
\end{equation}
which is  satisfied for $n =1 $ and it always satisfied for any 
$n > 1$. Since $\lambda_{-}$ is  negative for any $n$ 
$b(\eta) \sim (-\eta)^{\lambda_{-}}$ will be growing for any 
$n$. It is useful to look also at the cosmic time picture.
In cosmic time $b(t) \sim (-t)^{\zeta_{-}}$. Since 
$\dot{b}>0$ for any $n$ the solution is expanding 
also in cosmic time
\footnote{ Notice that the 
the over-dot denotes derivation with respect to cosmic time.
The kinematical conditions on the expansions or 
contraction of a given background geometry should be 
stated in cosmic rather than in conformal time.}.

In order to produce large scale magnetic fields when the 
internal dimensions shrink according to the Kasner solutions we have 
to require
\begin{equation}
n\lambda_{+} > 2.7. 
\label{c2}
\end{equation}
If we insert the explicit expression of $\lambda_{+}$ we can see that
Eq. (\ref{c2}) imposes the following inequality
\begin{equation}
\sqrt{\frac{3 n}{n + 2}} \geq 2.7
\end{equation}
which can never be satisfied for any $n$.
If the dynamics of the internal dimensions follows 
vacuum (Kasner) solutions, then, the requirement of generating 
large scale magnetic fields compatible with the cosmological bounds 
automatically selects an expanding dynamics of the internal 
dimensions.

Vacuum Kasner solutions can be generalized to the 
case of vanishing scalar potential. In this case  the equations of 
motion can
be always solved and  the power-law ansatz can still be made for the scale
 factors.
The solutions will be given by:
\begin{eqnarray}
&& (d -1) \sigma + n\lambda = 1
\nonumber\\
&& {\varphi '}^2 = 2\{ 1 - (d -1) \sigma^2 - n\lambda^2 + 2 \sigma\} 
\frac{1}{\eta^2}
\end{eqnarray}
Also in this case we can easily show that if the solution is contracting 
(in the internal manifold) large scale magnetic fields 
cannot be produced. Indeed if $b(\eta)$ contracts we will have that 
the allowed range of $n\lambda$ is given by $ 0<n \lambda <d$. If we take 
$d=3$ we see that $n \lambda < 3$. But according to our analysis 
$n\lambda > 2.7$. Therefore only a tiny slice of parameter space 
(i.e. $2.7 < n \lambda < 3$ ) could give 
a positive result. 

\subsection{Explicit examples}
Consider, as an example, the case $n=6$. In this case 
the internal manifold has six internal scale 
factors $b(\eta)$. As assumed in Eq. (\ref{metric}) 
both the internal and the external metrics will 
be maximally symmetric. Suppose that the internal 
dimensions are then expanding according to the 
Kasner solutions, namely $ b(\eta) \sim (-\eta)^{\lambda_{-}}$ and 
$a(\eta)\sim (-\eta)^{\sigma_{-}}$. 
If $n=6$ $\lambda_{-} = -1/4$ and $\sigma_{-} = 5/4$. 
The the background model given in Eq. (\ref{back}) will 
be in this case 
\begin{eqnarray}
&& a(\eta) = a_1 \biggl(-\frac{\eta}{\eta_1}\biggr)^{5/4},\,\,\,\,
   b(\eta) = b_1 \biggl(-\frac{\eta}{\eta_1}\biggr)^{-1/4},\,\,\,\,\,\,
   \,\,\,\, \eta< - \eta_1,
\nonumber\\
&& a(\eta) = a_1\biggl(\frac{ \eta + 2 \eta_1}{\eta_1}\biggr),\,\,\,\,\, 
   b(\eta) = b_1,\,\,\,\,\,\,\,\,\,\,\,\,\,\,\,\,\,\,\,\,\,\,\,\,\,\,\,\,
-\eta_1 \leq \eta \leq \eta_2,
\label{ba1}
\end{eqnarray}
where a sudden match to the radiation dominated epoch is 
assumed in $-\eta_1$. The internal scale factors grow from 
$\eta \rightarrow - \infty$ up to $-\eta_1$. 
The external scale factor contracts. This is 
because of the mathematical nature of vacuum Kasner 
solutions. 
Using Eqs. (\ref{vec1}) and (\ref{V})--(\ref{phi}) the
evolution of the mode functions during the 
multidimensional phase will be given by 
\begin{equation}
\phi_{k}''- \biggl[ k^2 - \frac{21}{16 \,\,\eta^2}\biggr] \phi_{k} =0,
\end{equation}
whose solution is given by Eq. (\ref{s1}) with $\nu = 5/4$.
At the time $-\eta_1$ the background will 
pass to radiation with a temperature $T_1\sim M_{4}$.
The produced large scale magnetic fields will 
evolve according to flux conservation and 
the ratio between the magnetic and radiation 
energy density is roughly constant. Then, from Eqs. 
(\ref{sp0})--(\ref{r1}), $r(\omega)$ can be 
explicitly written as 
\begin{equation}
r(\omega) = \biggl( \frac{M_{4}}{M_{P}}\biggr)^{4} \,\,
\frac{{\cal C}(5/4)}{\pi^2} 
\biggl(\frac{\omega}{\omega_1}\biggr)^{1/2}, \,\,\, 
\omega_{\rm dec}<\omega < \omega_1 
\end{equation}
where 
$\omega_1 \sim 10^{11}\,\,(M_{4}/M_{P})$ Hz and 
$\omega_{\rm dec} \sim 10^{-16}$ Hz are both evaluated 
at the present time and where ${\cal C}(5/4)/\pi^2 \sim 
1.86\times 10^{-3}$ [see Eq. (\ref{C})].
The value of $r(\omega)$ at the scale relevant 
for magnetogenesis (i.e. $\omega_{G} \sim 10^{-14}$ Hz)
is given by 
\begin{equation}
r(\omega_{G}) =1.86 \times 10^{-15.5} \,\,\,\,
 \biggl(\frac{M_{4}}{M_{P}}\biggr)^{3.5}. 
\label{exa}
\end{equation}
As discussed in Eq. (\ref{d1}), in order 
to have a successful magnetogenesis 
$r(\omega_{G}) > 10^{-33}$ 
should be required. Imposing this 
requirement it turns out that 
from Eq. (\ref{exa} 
\begin{equation}
\biggl(\frac{M_4}{M_{P}}\biggr) \geq 10^{-5}.
\label{sc1}
\end{equation}
If a less conservative magnetogenesis 
requirement is assumed (i.e. $r(\omega_{G}) \geq 
10^{-50}$) it can be checked that 
\begin{equation}
\biggl(\frac{M_4}{M_{P}}\biggr) \geq 10^{-10}
\label{sc2}
\end{equation}
should be demanded. 
If $M_4 \sim 10^{12} $ TeV both 
Eqs. (\ref{sc1})--(\ref{sc2}) 
are satisfied and $\omega_1 \sim 10^{7}$ Hz.
Moreover, $r(\omega_{\rm dec}) \sim 10^{-30}$
with negligible impact on the CMB.
If $M_{4} \sim 10^{7} $ TeV large scale 
magnetic fields can be generated only if 
a higher dynamo amplification 
rate is assumed, 
so that Eq. (\ref{sc2}) can be applied. 
In this case $\omega_1 \sim 100$ Hz.

\renewcommand{\theequation}{4.\arabic{equation}}
\setcounter{equation}{0}
\section{Discussion and Conclusions} 

In this paper the possible generation 
of large scale magnetic fields in the case 
of dynamical internal dimensions has been considered. 
The time dependence of 
internal dimensions (a generic 
ingredient of Kaluza-Klein models and of
superstring theories) is severely constrained 
by our knowledge of BBN. 
The fact that internal dimensions 
could not be dynamical after the 
Universe was old of approximately one second 
does not exclude that they could have been 
dynamical for temperatures 
larger that $1$ MeV. 

Since internal dimensions naturally break the 
conformal invariance of the evolution 
equations of Abelian gauge fields propagating 
in the four-dimensional world, electromagnetic fluctuations 
can be amplified.
The typical amplitude and scale of the  produced
magnetic fields has been investigated under the 
assumption  that the internal dimensions were dynamically 
evolving 
prior to the radiation dominated epoch.

In this context the problem of generating a seed 
field for the galactic dynamo action has 
been studied. The problem of magnetogenesis 
amounts to satisfying  the numerical requirements 
necessary for  a successful dynamo action together 
with all the bounds imposed by the standard model of 
cosmological evolution.
If the conformal 
time evolution of the internal dimensions is monotonic
and if the multidimensional phase is 
suddenly matched to radiation, then magnetogenesis
is possible provided the dimensions belonging to the internal 
manifold are expanding. If the internal dimensions 
are contracting magnetogenesis is not possible. 

The considerations reported in this 
paper should emerge as a part of a complete 
cosmological scenario. This problem has been 
only partially addressed in our considerations and 
further work is certainly needed. However, the 
idea of connecting large scale magnetic fields
with the dynamics of internal dimensions seems to 
be of some interest.

\section*{Acknowledgments}
The author wishes to thank M. E. Shaposhnikov 
for useful discussions and valuable exchanges of ideas.

\end{document}